# A Compact and Lightweight Fibered Photometer for the PicSat Mission

M. Nowak, S. Lacour, V. Lapeyrère, L. David, A. Crouzier, G. Schworer, P. Perrot, S. Rayane
LESIA, Observatoire de Paris, PSL Research University, CNRS,
Sorbonne Universités, UPMC Univ. Paris 06, Univ.
5 place Jules Jansen, 92195 Meudon;
mathias.nowak@obspm.fr

**ABSTRACT**

PicSat is a nanosatellite developed to observe the transit of the giant planet β Pictoris, expected in late 2017. Its science objectives are: the observation of the transit of the giant planet's Hill sphere, the detection of exocomets in the system, and the fine monitoring of the circumstellar disk inhomogeneities. To answer these objectives without exceeding the possibilities of a 3-unit Cubesat in termes of mass and power budget, a small but ambitious 2 kg opto-mechanical payload was designed. The instrument, specifically made for high precision photometry, uses a 3.5 cm effective aperture telescope which injects the light in a single-mode optical fiber linked to an avalanche photodioode. To ensure the stability of the light injection in the fiber, a fine pointing system based on a two-axis piezoelectric actuation system, is used. This system will achieve a sub-arcsecond precision, and ensure that an overall photometric precision of at least 200 ppm/hr can be reached.

**INTRODUCTION**

Beta Pictoris is a very famous star among astronomers, especially within the exoplanet science community. The discovery of a debris disk around this young (< 12 Myr), close-by (~ 20 pc), and thus very bright (Mv = 3.86) star in 1984 has raised the interest of the community for its circumstellar environment.

In 2003, observations performed with the Very Large Telescope (VLT) and the NaCo instrument, revealed the existence of a planet orbiting around this star[7]. Six years later, new observations with the same instrument showed that the planet Beta Pictoris b was a giant (6 to 12 Jupiter mass), orbiting at 8 to 15 Astronomical Units (AU) of its parent star. The orbital parameters were since refined by other observations[1], giving a semi-major axis of 8 to 9 AU, and an eccentricity < 0.26. But because all studies of this object have been made via direct imaging, the physical diameter of the planet, and thus its density and formation history remain unknown.

The discovery of Beta Pictoris b occurred in 2003, but the first hints of its existence were actually found before the VLT observation. In 1981, while Beta Pic was being observed as a reference star (!), a photometric event was detected in its light curve. Lecavelier et al.[3] suggested that this could be the sign of the presence of a giant planet transiting in front of the star. In recent works, Lecavelier at al.[4] showed that the orbit of Beta Pictoris b was indeed consistent with a close to 90 degrees inclination, and with a photometric transit in 1981. If true, this hypothesis also predicts a new transit in 2017 or 2018.

Up to this date, the best estimates of the orbital parameters suggest that the planet itself is not transiting its star, but that its Hill Sphere (its area of gravitational influence) does. Photometric variations are expected between March 2017 and March 2018, and will provide a unique opportunity to probe the planetary environment of Beta Pic b, and to better understand the disk-planet interactions.

Up to this date (June 2017), no sign of transiting material has been observed. But Beta Pictoris is a star only visible from the Southern hemisphere, and hidden by the Sun during summer. So despite the efforts of the astronomical community, and the number of projects set up to cover the event, detecting light variations from the ground will be difficult. A couple of observations are planned using space telescopes (BRITE, or the HST), but they have their limitations in terms of observing time.

The PicSat mission, expected to launch in September 2017 will monitor the last half of this extremely rare event, and will provide a continuous and homogeneous photometric coverage of the closest approach, half Hill radius egress and final egress (which, together with the ingress and the half Hill radius ingress are the 5 times when the transit of material around the planet is most likely).

To do so, the mission must embark an instrument capable of performing continuous high-precision photometry over an extended period of time. The mission format (a 3 units CubeSat, measuring 30 cm x 10 cm x 10 cm), and its weight limitation (< 5 kg) puts strong constraints on the design of this instrument. The PicSat fibered photometer has been developed to answer this challenge.

**SCIENCE OBJECTIVES**

*Transit of Beta Pic b's Hill sphere*

The main and most important science objective of the PicSat mission is the monitoring of the transit of Beta Pictoris b's Hill sphere in front of its star, the measurement of its radius, and the detection of transiting material. Based on the light curve obtained in 1981, is can be estimated that this requires a photometric level of precision of at least $10^{-3}$ (7-sigma) to detect the event. A level of $10^{-4}$ would be better suited to characterize the transit.

*Studying exocomets in the Beta Pic system*

The giant planet Beta Pic b is not the only astronomical body which is expected to transit in front of the star. A number of smaller objects, identified as exocomets[2]. But to this date, these exocomets have only been seen by transit spectrometry, where absorption features are detected at certain wavelengths, revealing the gas tail of the comet. Lecavelier et al.[5] have shown, using a dedicated model of transiting exocomets, that these objects could be observed in large band with a photometric level of precision of $10^{-4}$/hr. A secondary objective of the PicSat mission is to detect these exocomets in the visible band, thus opening up the possibility of studying the dust tail of these objects.

*Disk inhomegeneities*

The PicSat mission and its fibered instrument will provide a unique dataset of photometric measurements with the rare combination of precision, homogeneity, and long duration that only a space mission can achieve. This will help to study and hopefully to better understand the structure of this young debris disk, in which planet(s) have very recently been formed.

*Single mode fiber in space*

Last but not least, the PicSat instrument has also been designed with a technical objective in mind: demonstrating our ability to properly inject star light into a single mode fiber in space. Single mode fibers are now broadly used to collect an d guide light in astronomical instruments. They are especially useful for interferometric instrumentation, where their propensity to retain the coherence of the light is especially useful. But because of the very small diameter of the core of these fibers (3 microns), proper light injection is difficult, and requires advanced guiding/tracking techniques. The PicSat fibered photometer will prove our ability to manage this injection, and pave the way to future projects, like the FIRST-S nanosatellite[6].

**MISSION OVERVIEW**

*Concept*

The PicSat platform, which will embark the PicSat fibered photometer, is a 3 unit CubeSat, specifically designed for high precision transit photometry. An overview of the general design is given in Figure 1. One of the 3 units is dedicated to the On-Board Computer (OBC), power system, and communication board. The second unit contains the Attitude Determination and Control System (ADCS), and the payload electronics. The last unit is reserved to the payload opto-mechanics, and also hosts the star tracker (part of the ADCS).

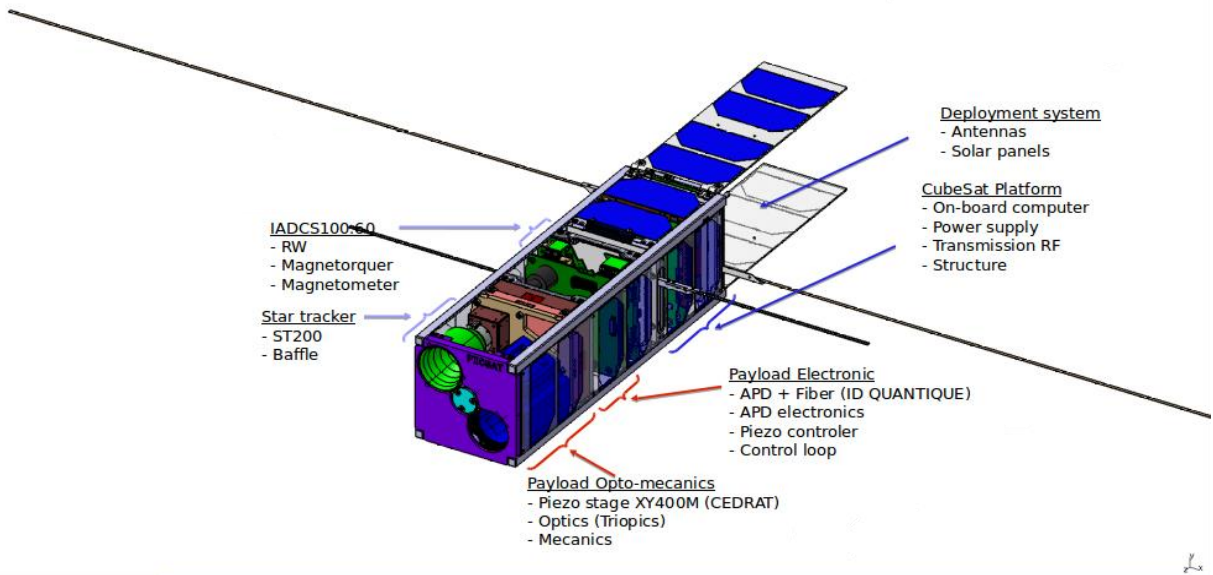

**Figure 1: Overview of the PicSat platform**

The satellite is currently scheduled for launch in September 2017. The initial mission duration is 1 year.

*Orbit*

The orbit of the satellite is mainly constrained by Sun exposition (for power generation), observability of Beta Pictoris (RA 05h 47min 17s, DEC = −51deg 03' 59"), ground station visibility, and launch opportunities. A large part of the launch available launch opportunities concern polar Sun-Synchronous Orbit (SSO), very well-suited for Earth observation. Such orbits combine a constant illumination angle over the year, and good visibility of Beta Pictoris (> 60 % of the time, depending on the altitude), making them perfectly suited to the PicSat mission.

We studied different possibilities, with altitudes ranging from 500 km to 650 km, and found that in all cases, the total fraction of Beta Pic visibility (≥ 63%), and Sun exposition (≥ 75%) was compatible with our objectives.

**PICSAT FIBERED PHOTOMETER DESIGN**

*Concept and error budget*

To tackle its science objective, the PicSat mission requires an instrument capable of simultaneously demonstrating the use of a single mode fiber in space, and achieving high-precision photometry. This lead to the design of a "fibered photometer", in which, contrary to all usual photometers, the 2 dimensional detector array is replaced by a Single Pixel Avalanche Diode (SPAD). The concept is shown in Figure 2.

The light coming from the stars is collected by a small optical telescope (effective diameter is 3.5 cm), injected in a Single Mode Fiber (SMF) placed in the focal plane of the telescope, and brought to the SPAD for photon counting. The fiber has a core diameter of 3 microns only, similar to the diameter of the star image on the focal plane of the instrument (Point-Spread Function, PSF). Thus, to properly inject light into this fiber, a specific tracking mechanism is required, to ensure that the small fiber stays centered on the star, even in the presence of guidance system errors and/or jitter.

To track the star in the focal plane of the telescope, the fiber is mounted on a two-axis piezoelectric actuator. This actuator can move the fiber in the focal plane, over a 450 microns x 450 microns area (about 1' x 1' in terms of field of view). The ADCS performance level ensures that the satellite is able to point Beta Pictoris with a sufficient precision so that its image will actually falls into the accessible range of the fiber. The piezo actuator will then move the fiber to "scan" the entire field, find the star, and track it.

To reach a level of precision of 100 ppm/hr on the photometry of the star, an excellent tracking is not enough. Some variations of the PSF of the instrument are expected (mainly due to thermal stress of the optics), and will result in instrumental variations of the photometry. To correct these, it is necessary that the "tracking" algorithm also regularly "scans" the PSF, and estimates some of its most critical parameters (mainly its size in two orthogonal directions). To do so, the piezo actuator will constantly modulate the position of the fiber around the central position of the star. Different modulation patterns are still being studied, but so far, the pattern given in Figure 3 seems to give good results. The modulation will run at 100 Hz, with integrations of 1 ms (photometric data point are acquired at 1 kHz, and the modulation pattern contains 10 points).

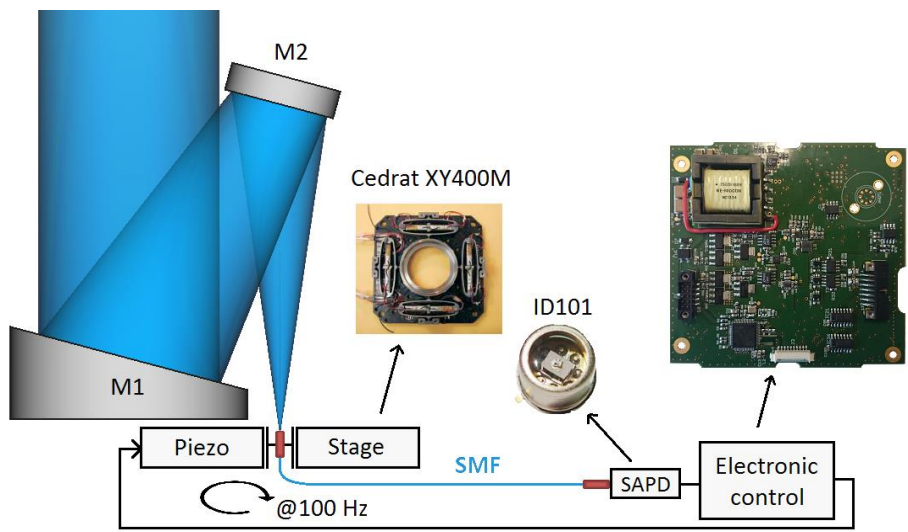

**Figure 2: PicSat Fibered Photometer concept**

All payload activities are controlled and managed by a dedicated electronic board, which embeds a 72 MHz STM32F303 microchip.

An error budget for the photometric precision of this instrument is given in Table 1. The use of a Single Mode Fiber, with its very small acceptance angle, drastically reduces the scattered light noise. The use of a photon counting detector (the SPAD) removes any source of readout noise. The main sources of noise are then the injection stability (related to the tracking precision), the thermal regulation of the detector, and quantum photon noise.

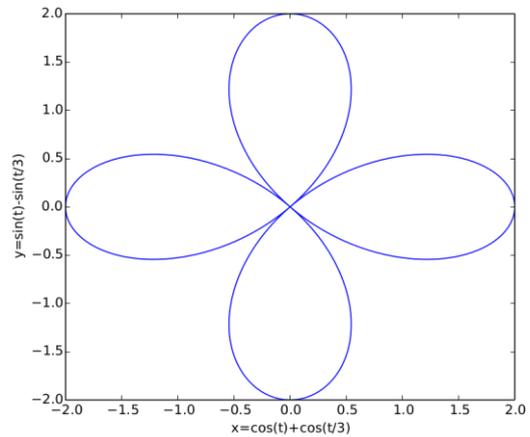

**Figure 3: One of the modulation pattern for scanning the PSF of the star**

*Opto-mechanical design*

For proper injection into the fiber, the aperture ratio of the telescope must match the one of the fiber. Thus, the aperture ratio of the PicSat telescope is constrained to F/D = 4. The telescope must also fit into a single CubeSat unit. A compact 30° off-axis Newtonian design was selected, with an effective diameter of 3.7 cm, and a focal length of 14.8 cm. The primary off-axis parabola used is oversized (50 mm diameter) to ensure optimum optical quality on the edges, and made of pure aluminum. The secondary mirror is a plane mirror, of 22 mm diameter. The overall optical design is shown in Figure 4.

**Table 1: Photometric error budget**

| Noise source | Assumption | Error (ppm/hr) | Comment |
|---|---|---|---|
| Photon noise | Mv = 3.86 | 60 | Physical limit |
| Readout noise | No readout noise | 0 | |
| Scattered light | 150 e-/s | 5 | Moon and Earth. Filtered by SMF |
| SPAD bias voltage stability | 150 μV | 20 | By design |
| Thermal stability | Regulated and corrected to 0.01 °C | 40 | By design and with data reduction |
| Injection stability | 5% at 100 Hz | 80 | From simulations |
| Total error | | 110 | Square-root of squared errors |

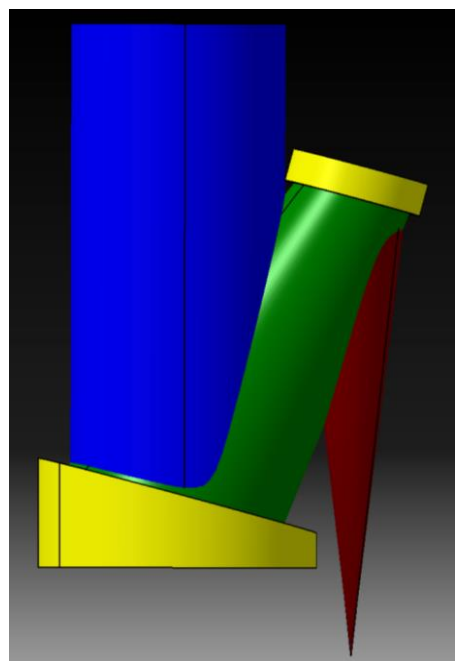

**Figure 4: Optical design of the Fibered Photometer**

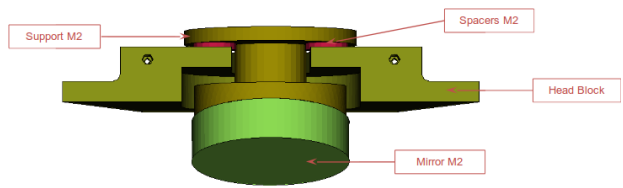

**Figure 5: A 3D view of the secondary mirror support.**

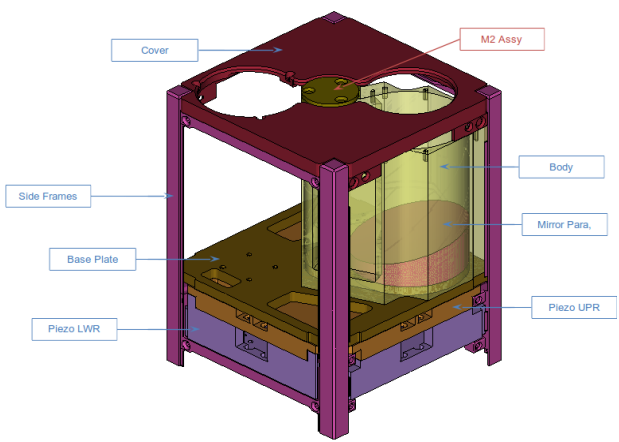

**Figure 6: Mechanical assembly of the PicSat Fibered Photometer, integrated into a standard CubeSat unit.**

The secondary mirror is mounted on a specific mechanical assembly which allows the fine tuning of its alignment with respect to the primary parabola thanks to the use of 3 peelable washers. A 3d rendering of this assembly is given in Figure 5. This assembly is mounted on top of the optical tube. The general mechanical assembly, integrated into a standard CubeSat unit is shown in Figure 6. The whole structure is made out of aluminum, so that thermal variations are expected to produce homogeneous deformations, minimizing their impact on optical quality of the system.

The fiber is mounted on the two-axis piezo actuator, and positioned in the focal plane of the telescope. The piezo actuator is based on space qualified components, and made by CEDRAT Technologies (Grenoble, France).

A space is left on the baseplate, next to the body of the telescope, to host the star tracker (part of the ADCS). This will ensure a rigid link between the two optical systems, and will help to maintain the relative position of their line-of-sights. This is of prime importance for the two-stage tracking system (ADCS + payload piezoelectric system).

*Electronics*

A dedicated electronic has been designed and realized for the PicSat Fibered Photometer. This board includes all the required electronics to manage data acquisition, communications to and from the payload, and to run the tracking algorithm. Typical power consumption is 0.2 W in "Standby mode", and 2 W in "Science mode".

The electronic board embeds a 72 MHz STM32F303 microchip. This component, whereas not space qualified, combine the necessary computing performance to run a Kalman filter based algorithm at high frequency (1 kHz) with a low power consumption profile.

The two-axis piezo actuator is driven by 2 Digital-to-Analog Converters (DACs), and a high voltage (150 V) source. The typical displacement step is 0.01 micron, and the total accessible range is 450 microns on both axes.

The SPAD is connected to a 5 V power supply and to a -25 V bias voltage source which are regulated by the board itself to ensure an excellent level of stability (100μV), as this is critical for the overall photometric precision. The output pin of the SPAD (electron cascade signal) is connected to a 4b-bit hardware counter. This counter is activated/deactivated by a 3.3 V signal outputted by one of the hardware timers of the STM microchip. This ensures a proper timing of the integration cycles.

The SPAD retained for the PicSat Fibered Photometer has a built-in Thermo-Electric Cooling (TEC) system, which can be used to regulate the temperature of its single pixel. The TEC is managed by a dedicated TEC control system, on the payload board. A simple DAC is used to set the temperature set-point (between 2°C and 16°C). The system provides good thermal stability (down to 0.1°C) over extended conditions.

Communications in both reception and transmission is achieved through a standard UART/RS422 4-wire interface.

**TRACKING ALGORITHM**

As stated above, one of the main role of the payload board of the PicSat Fibered Photometer to control 2-axis piezo actuator on which the SMF is mounted, to correct for pointing vibrations induced by the ADCS. The payload board has also in charge the initial scan of the focal plane of the telescope at first acquisition, or when the tracking is lost.

The algorithm used in the PicSat Fibered Photometer is based on Kalman filtering. To describe the state of the system, a standard position-velocity-acceleration representation could be used. However, in our case, the acceleration originates in the jitter created by the ADCS, and must be considered as a source noise. We thus use a simple position-velocity representation, in which we denote X the state of the system, where:

$$X = (x, y, \partial_t x, \partial_t y)^T \quad (1)$$

x and y being the position of the star in the focal plane, on both axes.

The evolution of the system between times $t_n$ and $t_{n+1}$ can then be written:

$$X_{n+1} = \begin{pmatrix} 1 & 0 & dt & 0 \\ 0 & 1 & 0 & dt \\ 0 & 0 & 1 & 0 \\ 0 & 0 & 0 & 1 \end{pmatrix} X_n + \begin{pmatrix} dt^2/2 & 0 \\ 0 & dt^2/2 \\ dt & 0 \\ 0 & dt \end{pmatrix} \times \begin{pmatrix} a_x \\ a_y \end{pmatrix} \quad (2)$$

where $a_x$ and $a_y$ are two independent and centered random variables. For the sake of simplicity, we suppose that these variables are gaussian, and we denote V the variance covariance matrix of the vector ($a_x$ $a_y$). In these conditions, if, at any given time t, the state $X_{n-1}$ of the system is known with some uncertainty characterized by its variance-covariance matrix $P_{n-1}$, then Equation 2 leads to the *a priori* estimate:

$$X_n^{ap} = AX_n \quad (3)$$

$$P_n^{ap} = AP_{n-1}A^T + QVQ^T \quad (4)$$

where A is the left matrix of the right-hand part of Equation 2, and Q the right matrix of the right-hand part.

Using Equations 3 and 4, it is possible to predict the evolution of the position of the star (the first two elements of X) over time. However, without any other information, this position will evolve at constant velocity. This should not be a surprise, as we used a position-velocity model. To have a reliable estimate of the position of the star over time, it is necessary to also include in the model the measurements obtained with the SPAD. This is easily done using the Kalman filter formalism.

Suppose that we have a set of m measurements of the flux injected in the fiber $c_{n-m+1}$, …, $c_n$ obtained over one modulation pattern, at times $t_{n-m+1}$, …, $t_n$, and positions $r_{n-m+1}$, …, $r_n$. An estimate of the position of the centroid can be obtained by computing the barycenter of the $r_k$, weighted by the $c_k$:

$$r_{n-m/2} = \frac{\sum_{k=n-m+1}^{n} c_k r_k}{\sum_{k=n-m+1}^{n} c_k} \quad (5)$$

Then, if we suppose that this estimate is unbiased and gaussian, with variance-covariance matrix $R_{n-m/2}$, the Kalman theory predicts the best *a posteriori* estimate of X is given by:

$$X_n = X_n^{ap} + K_n(r - CX_n^{ap}) \quad (6)$$

$$P_n = (I - K_n C)P_n^{ap} \quad (7)$$

where $K_n$ is the Kalman gain:

$$K_n = P_n^{ap} C^T (CP_n^{ap} C^T + R_n)^{-1} \quad (8)$$

and C is the "measurement prediction" matrix, used to predict the "measurement" $r_{n-m/2}$ from the state $X_n$:

$$C = \begin{pmatrix} 1 & 0 & m/2 \times dt & 0 \\ 0 & 1 & 0 & m/2 \times dt \end{pmatrix} \quad (9)$$

Using Equations (1) to (9), it is possible to iterate after each new 1 ms acquisition using the past few points (typically the past 10 points) to compute the best estimate of the position of the star in the focal plane, and recenter the modulation pattern on it for the next acquisition.

Using a Matlab/Simuling model which includes a simple model of ADCS errors (white noise between 0 to 10 Hz, with a very pessimistic 20 microns rms value, i.e. 30" precision only at 1 sigma level), and a complete model of the optical system (PSF of the instrument + injection into the fiber), we have shown that this algorithm can indeed lead to excellent tracking precision, with a residual error as low as 0.3 micron rms (see Figure 7).

**ENVIRONMENTAL TESTS**

Many of the elements of the Picsat Fibered Photometer (especially the electronics) are off-the-shelf industrial component, which are not space-qualified. To ensure that these components would net fail under the harsh condition of the launch/space environments, we conducted vibration and thermal-vacuum test campaigns.

*Vibrations*

The full instrument (opto-mechanical + electronics) was integrated into a 3 unit CubeSat structure, and tested in vibrations. Quasi-static tests (5 – 100 Hz, at 9.7 g), sinusoidal vibrations (5.0 – 8.0 Hz at 10 mm amplitude, and 8 – 100 Hz at 4.5 g), and random vibrations (6.7 g during 120 s) were performed. The instrument systems showed no obvious sign of damage, and all subsystems responded well to a detail health-check. In particular, a complete scan of the optical response (instrumental PSF) before and after the tests proved that the optical alignment withstood the vibrations.

*Thermal vacuum*

The full instrument, also integrated in a 3 units CubeSat structure, was placed in a thermal vacuum chamber, and was cycled between -20°C and +30°C. The electronics responded well under any of these conditions, but the secondary mirror, initially made of Zerodur glass ceramic and aluminum coating was damaged. The Zerodur-aluminum combination is not known to be resilient to the space environment[8], and the mirror is currently being replaced by a space-qualified component. Otherwise, all other subsystems (including the primary parabola, the electronic board, the SPAD, the thermal regulation system, etc.) withstood the test very well, and we are now confident that the photometric performance of the PicSat Fibered Photometer will not be impacted by the space environment.

**CONCLUSION**

This paper presented the PicSat Fibered Photometer developed to observe the transit of Beta Pictoris b in late 2017.

By combining an innovative approach to photometric observations using a single mode fiber and a single pixel avalanche detector, together with a more standard optical design (an off-axis Newtonian telescope), the PicSat Fibered Photometer achieve high-precision photometry while keeping an overall size, mass and power consumption extremely low. The final instrument fits in a standard CubeSat unit (10 cm x 10 cm x 10 cm), weights 1.3 kg only, and is capable of achieving 100 ppm/hr photometric precision on stars with Mv < 5.

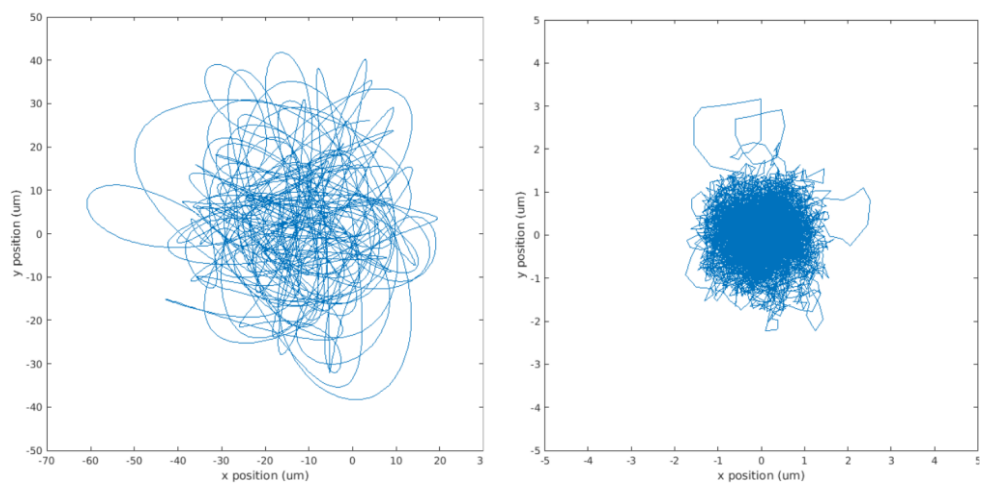

**Figure 7: Result of a Matlab/Simulink simulation presenting the position of the star with respect to the SMF when the tracking algorithm is off (left panel) and on (right panel).**

The instrument has passed vibrations and thermal vacuum tests with success, demonstrating its ability to reach its objectives even in the harsh conditions of space.

The flight model will be integrated in the PicSat platform in July 2017, and is scheduled for launch in September 2017. High-precision photometric observations of Beta Pictoris will be available soon!

*Acknowledgments*

This work was supported by the European Research Council (ERC-STG-639248), the MERAC foundation (Mobilising European reserach in Astrophysics and Cosmology), and the Labex ESEP (Exploration Spatiale des Environnements Planetaires)